\documentclass[doublecol,figures]{epl2}

\usepackage{amsfonts}

\title{Magneto-optical imaging of magnetic deflagration\\ in Mn$\bm{_{12}}$-Acetate}

\author{D. Villuendas\inst{1} \and D. Gheorghe\inst{2} \and A. Hern\'andez-M\'inguez\inst{1} \and F. Maci\`a\inst{1} \and J. M. Hernandez\inst{1} \and\\ J. Tejada\inst{1} \and R. J. Wijngaarden\inst{2}}
\shortauthor{D. Villuendas \emph{et al.}}

\institute{
    \inst{1} Departament de F\'isica Fonamental, Facultat de F\'isica, Universitat de Barcelona - Avda. Diagonal 647,\\ Planta 4, Edifici Nou, 08028 Barcelona, Spain, EU\\
    \inst{2} Faculty of Sciences, Division of Physics and Astronomy, Vrije Universiteit - De Boelelaan 1081,\\ 1081 HV Amsterdam, The Netherlands, EU
}

\pacs{75.50.Xx}{Molecular magnets}
\pacs{75.60.Jk}{Magnetization reversal mechanisms}
\pacs{85.70.Sq}{Magneto-optical devices}

\abstract{
For the first time, the morphology and dynamics of spin
avalanches in Mn$_{12}$-Acetate crystals using magneto-optical
imaging has been explored. We observe an inhomogeneous relaxation of the
magnetization, the spins reversing first at one edge of the
crystal and a few milliseconds later at the other end. Our data fit well with the theory of magnetic deflagration, demonstrating that
very slow deflagration rates can be obtained, which makes new types of
experiments possible.}

\begin{document}

\maketitle

Synthesized in 1980, Mn$_{12}$-Acetate is a crystal composed of a
large number (typically of the order of $10^{17}$) of
[Mn$_{12}$O$_{12}$(CH$_3$COO)$_{16}$$\cdot$(H$_2$O)$_4$]$\cdot$2CH$_3$COOH$\cdot$4H$_2$O
molecules, each with a large spin $S = 10$ $\mu_B$ \cite{Lis}. Its
mono-disperse structure, strong spin anisotropy as well as the
hysteretic behavior at low temperatures \cite{Sessoli} have
attracted a lot of interest. From a fundamental point of view,
Mn$_{12}$-Acetate offers an unique playground to study phenomena
at the frontier between classical and quantum mechanics, whereas
from the point of view of applications, it offers an interesting
perspective towards 3D high-density magnetic storage devices
\cite{Srajer}. These possibilities motivated many studies on the
magnetic properties of these molecular crystals (see refs.
\cite{revFriedman,revGatteschi,revdelBarco} for reviews). The
strong uniaxial anisotropy accounts for the doubly-degenerate
potential well, shown in fig.1(a). Each well is characterized by a
discrete distribution of energetic levels, corresponding to
different projections $m$ ($m = \pm10, \pm9,...0$) of the total
spin along the easy anisotropy axis ($c$-axis). In such a system,
the relaxation of the magnetization can proceed via three
mechanisms: thermal relaxation, quantum tunnelling and avalanches.
Thermal relaxation occurs when the thermal excitations are strong
enough to promote the spins over the potential barrier. When this
condition is not fulfilled, the effect of
\begin{figure}[ht!]
\center
\onefigure[width=0.62\columnwidth]{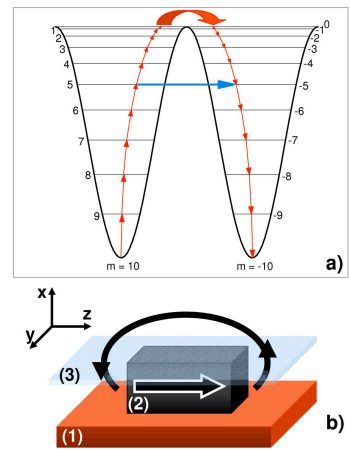}
\caption{(Color online) a) Doubly degenerate potential well of a
Mn$_{12}$-Acetate crystal in zero applied magnetic field. The
possible relaxation mechanisms are schematically indicated by the
arrows: thermal relaxation (red arrows), in which the thermal
energy promotes the spins above the potential barrier in the
adjacent well, and quantum tunnelling (blue arrow), in which the
spins thread through the energy barrier in the adjacent well. b)
Schematic representation of the experimental configuration: (1)
Cernox resistor, (2) Mn$_{12}$-Acetate crystal, (3)
magneto-optically active layer (indicator). \label{Fig.1}}
\end{figure}
 the thermal energy is
simply to populate the upper levels of the potential well, thus
increasing the propensity for spin tunnelling through the
anisotropy barrier \cite{LibroTejada}. Quantum tunnelling effects
appear irrespective of temperature, as steps in the hysteresis
curve which occur systematically at well defined externally
applied magnetic fields, the so-called resonant fields $H_{res}$
\cite{QTM,JMepl,JMPRB}. At $H_{res}$ some energy levels in both
wells are at equal energy.

Thermally induced relaxation processes are characterized by a long
relaxation time $\tau$, typically of the order of a few hundred
ms. Much shorter relaxation times, less than 1 ms
\cite{Paulsen,Suzuki,AHMprl,AHMjm3,McHugh,AHMeplMnCr}, have been
observed experimentally. Since in such cases the relaxation of the
magnetization is typically accompanied by a significant heat
release this effect is attributed to a thermal runaway or
avalanche. It was observed that this spin reversal (avalanche)
does not occur simultaneously for all the spins in the sample, but
follows a domino effect, with spin reversal nucleation at one edge
of the sample and subsequent propagation through the crystal over
a narrow spin reversal front \cite{Suzuki}. This type of
relaxation is known as magnetic deflagration, due to its
parallelism with the propagation of a chemical combustion along a
burning substance. In the magnetic case, the role of the chemical
energy is played by the Zeeman energy and the "ashes" are the
spins that have already relaxed. The amount of heat released by
the thermal runaway depends significantly on the experimental
parameters (temperature, sweep rate of the magnetic
field)\cite{delBarco}. However, up to now, no correlation was made
between the external parameters, the excitation energy and the
morphology of the spin reversal fronts of the magnetic
deflagration. In this study we set on providing an answer to this
open question by using for the first time an optical technique:
magneto-optical imaging. This technique allows the direct
visualization of the overall magnetic field distribution at the
surface of the sample as a grey-scale intensity image, the value
of each grey level being proportional to the local value of the
magnetic field component perpendicular to the surface of the
sample \cite{Koblischka} (i.e. the $x$-axis in fig.1(b)). A
detailed description of our magneto-optical set-up can be found in
ref. \cite{Wijngaarden}. Our sample, a Mn$_{12}$-Acetate crystal
of $1.2\times0.5\times0.5$ mm$^3$, was mounted on a LakeShore
Cernox Thermometer, model CX-1050BR, which we used as a heater.
The sample was visualized using the Faraday effect in a 5 $\mu$m
Bi-substituted yttrium iron garnet (YIG) film with in-plane
anisotropy \cite{Dorosinskii} and a saturation field of 90 mT. The
magneto-optically active layer was 'glued' on top of the Mn$_{12}$
crystal with nanodecane (C$_{19}$H$_{40}$), 99$\%$ purity . A
schematic view of the experimental configuration is shown in
fig.1(b). The ensemble was mounted on a home-built optical insert
and placed in a commercial Oxford Instruments vector magnet. The
system allows the generation of magnetic fields up to $H_{max}=1$
T, in any orientation relative to the sample, by simultaneous use
of three superconducting coils. Since at low temperature an
external field of $1$ T is too small to ensure the saturation of
the magnetization, we prepared the initial state of the sample as
follows: we field cooled the crystal in an applied magnetic field
of $1$ T, oriented along the anisotropy axis ($z$-axis in fig.1
(b)), through the blocking temperature $T_B$ (above which thermal
relaxation is the dominant relaxation mechanism)
\cite{revFriedman}, down to the desired temperature $T$. Once $T$
was reached, we decreased the externally applied magnetic field
$H$ from $1$ T to a smaller value, indicated below, and
anti-parallel with respect to the magnetic moment of the sample.
Subsequently, voltage pulses of 10 V and various duration $t_p$
were applied to the Cernox resistor, using a Wavetek pulse
generator. We performed two sets of experiments, which we will
henceforth refer to as \emph{experiment 1} ($T = 1.5$ K, $H =
-0.1$ T, $t_p =10$ ms) and \emph{experiment 2} ($T = 1.6$ K, $H =
-0.45$ T, $t_p = [0.01, 10]$ ms), in which the initial state was
prepared using different values for the external parameters, as
indicated in the brackets. The magnetization reversal was filmed
using a Teli CCD camera (model CS8320C) with an active area of
$752\times582$ pixels and a recording speed of 50 fields/second.
The relatively low temporal resolution of our experiments
motivated the use of small magnetic fields $H$, since a smaller
$H$ is known to lead to slower relaxation \cite{QMDTheory}.

\begin{figure*}[ht!]
\begin{center}
\includegraphics*[scale=0.7, angle=270,viewport=40 0 350 700]{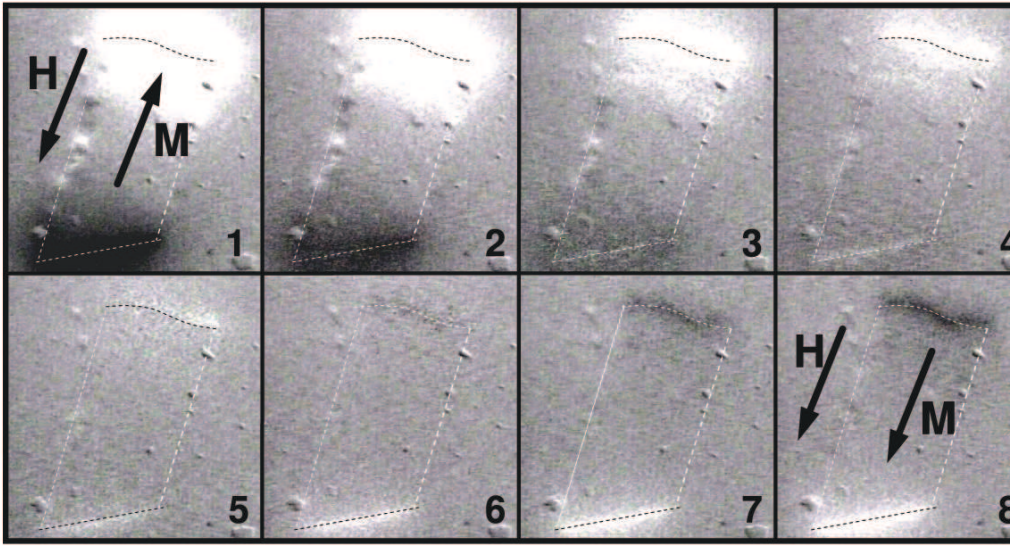}
\caption{Images showing the time evolution of the in-plane
magnetic moment of the sample obtained in \emph{experiment 1}. The
image in fig.2(1) corresponds to the initial state, prior to the
application of the heat pulse. All subsequent pictures were taken
after the heat pulse. The time span between consecutive frames is
20 ms. The magnetization reversal nucleates at the lower edge of
the sample and propagates, along the anisotropy axis,towards the
upper edge of the crystal. The intensity of the gray-levels in the
images is proportional to the intensity of the local magnetic
field. The arrows indicate the orientations of the magnetic moment
$M$ and the externally applied field $H$. The contour of the
sample is indicated by the dashed lines.\label{Fig.2}}
\end{center}
\end{figure*}

The results of \emph{experiment 1} are shown in fig.2. Fig.2(1)
corresponds to the initial state, prior to the application of the
heat pulse. The sample has a strong in-plane magnetic moment which
appears as the contrasting bright and dark regions at the upper
and lower edges of the crystal. The bright region corresponds to a
high positive magnetic field whereas the dark region to a high
negative magnetic field. After the heat pulse (which yields an
energy of 50 $\mu$J (see footnote ~\footnote{Note: over the range of temperatures at which we performed our
experiments, the resistance of the Cernox Thermometer is R = 20
k$\Omega$, as specified by the manufacturer. The Joules energy
dissipated due to the applied voltage pulses $V$ was calculated
using $E_J = V^2t_p/R$.}) is applied, a slow
decrease of the contrast at the edges of the sample is initially
observed (fig.2(2)). This corresponds to a decrease of the local
magnetic fields, hence to a reduction of the in-plane magnetic
moment of the sample. Subsequently, bright regions become clearly
visible on the dark background at the lower edge of the sample,
fig.2(3), where gradually, the field changes sign, fig.2(4).
Clearly, the deflagration and its associated sign reversal of
magnetization has started at the lower edge. The upper edge is
still unaffected and remains in its original magnetization
direction, see fig.2(5). Eventually, also the local magnetic field
at the upper edge changes sign. The magnetization reversal
gradually continues, until the magnetic moment of the sample has
completely reversed, fig.2(6), (7) and (8). It can be noticed, by
comparing fig.2(1) and fig.2(8), that the strength of stray field
at the edges of the sample is much stronger prior to the
application of the heat pulse. This is due to the fact that the
magnetic moment relaxes toward the equilibrium state, which, for
the values of $H$ used in our experiments, is lower than the
saturation magnetization: fig.2(1) is the magnetization induced by
the external field of 1 T and fig.2(8) is the magnetization
induced by 0.1 T only.

The relaxation process observed in our experiments is similar to
the one reported in ref. \cite{Suzuki} in the sense that the
magnetization reversal nucleates first at the edge of the sample.
Additionally, the relaxation starts systematically at the lower,
sharper edge of the sample in all our measurements. This agrees
well with the results reported in ref. \cite{AHMeplMnCr,Jaafar},
where it was observed that the deflagration front is ignited at
the point of the crystal where the local magnetic field is
highest. In our case, the demagnetizing field  $H_d$ is higher at
the sharper edge, which implies that irrespective of the sign of
the applied field $H$, the total magnetic field ($H + H_d$) at
this edge is highest. However, by contrast to the results of ref.
\cite{Suzuki}, no gradual propagation of the spin reversal, over a
narrow front along the width of the sample, was observed in our
experiments. Furthermore, the time span for complete relaxation is
larger in our experiment as compared to ref. \cite{Suzuki} and, at
first sight one could imagine that we are observing conventional
thermal relaxation, induced by the heater.

To try to understand the differences between our experimental
results and the data of ref. \cite{Suzuki}, we modified our
experimental conditions as detailed above for \emph{experiment 2}.
This was done in order to reduce the anisotropy barrier and probe
the avalanche nucleation at very small thermal excitations. The
time span $t_p$ over which the heat pulses were applied were
logarithmically increased from 0.01 ms to 10 ms. For very short
time pulses, $t_p = 0.01$ ms (equivalent energy 50 nJ), we
observed no relaxation of the magnetic moment, indicating that in
this case the excitation energy is too small to promote the
magnetization reversal. In all other experiments, corresponding to
$t_p=0.1$ ms (0.5 $\mu$J), $t_p=1$ ms (5 $\mu$J) and $t_p=10$ ms
(50 $\mu$J) a complete reversal of the magnetic moment is
observed, the relaxation mechanism following the same pattern as
detailed for \emph{experiment 1}. For each of these experiments we
calculated the variation of the local field intensity at the edges
of the crystal as a function of time. The results are shown in
fig.3. The data from all three experiments collapse to the same
curves, which can be fitted by an exponential law with $\tau=65.8$
ms. A relaxation time of $\tau =65.8$ ms implies through
Arrhenius' law, $\tau =\tau _{0}\exp {\frac{U(H)}{k_{B}T}}$, that
the temperature of the sample, after the heat pulse, rises to
$T=U/\ln (\tau /\tau _{0})\approx 4.5$ K (assuming that $U=59$ K
when $H_{z}=-0.45$ T \cite{JMepl}). This value is higher than the
blocking temperature $T_B\simeq3.5$ K (determined from DC magnetization measurements), hence thermal
relaxation plays a role. On the other hand, the mechanism cannot
be simple thermal relaxation, since in that case the relaxation
time would decrease with increasing $t_p$, in contrast to our
experimental observations.

To solve this apparent contradiction we propose the following
explanation: the externally applied heat pulse (provided that
$t_{p}>0.1$ ms) triggers a deflagration wave. During the
avalanche, the spin reversal is associated to an energy release in
the sample equal to the Zeeman energy $E_z =
\mu_0H$$\Delta$\emph{M}$_z$, where $H$ is the applied magnetic
field and $\Delta$\emph{M}$_z$ = \emph{Ng}$\mu_B\Delta$\emph{m} is
the change in the total magnetic moment of the sample due to the
spins that have changed their projection from $m = -10$ to $m =
10$ (implying $\Delta$\emph{m} = 20). The total number $N$ of
molecules in the sample can be easily evaluated using $N =
v\rho$$N_A/M$, where $N_A = 6.022\cdot10^{23}$/mol is the Avogadro
number, the volume $v$ of the crystal is $v = 1.2 \times 0.5
\times 0.5 = 0.3\cdot10^{-3}$ cm$^3$, the density $\rho = 1.84$
g/cm$^3$ and the molecular mass $M = 2060,3$ g/mol. Using $g = 2$,
the calculation yields $N = 1.6\cdot10^{17}$, $\Delta$\emph{M}$_z
 = 6.0\cdot10^{-5}$ J/T, which amounts to a Zeeman energy $E_z
\simeq 30 \mu$J. This energy is larger by a factor of 60 compared
to the excitation energy ($\sim 0.5 \mu$J) for which we already
observe avalanches in our experiments. Since even for larger
excitation pulses (up to $50 \mu$J) the complete relaxation of the
magnetic moment of the sample occurs over the same time interval,
it is plausible to assume that, for the range of excitation
energies used in our measurements, only a small fraction of the
energy provided by the Cernox thermometer is transferred to the
crystal. Hence, the increase in the temperature of the sample is
mainly due to the Zeeman contribution, which is the same in our
experiments, since the externally applied magnetic field (hence
the distance between the energetic levels in the crystal) is the
same. The heat produced by the deflagration wave is mostly
released to the environment, such that the temperature of the
crystal does not significantly exceed $T_B$. This explains why we clearly observe
the in-plane magnetic moment of the sample at any moment during
the deflagration.

The temperature of the front of reversing spins in our sample is
lower than in previously reported experiments,\cite{Suzuki,AHMprl}
which implies a higher value of $\tau$, hence a slower relaxation,
as we clearly observe in our measurements. We estimate that the
speed of the front of reversing spins is of the order of $v\approx
12$ mm/s, in contrast to some much higher values reported in the
literature.\cite{Suzuki,AHMprl,AHMjm3,McHugh}.

\begin{figure}
\includegraphics[width=\columnwidth]{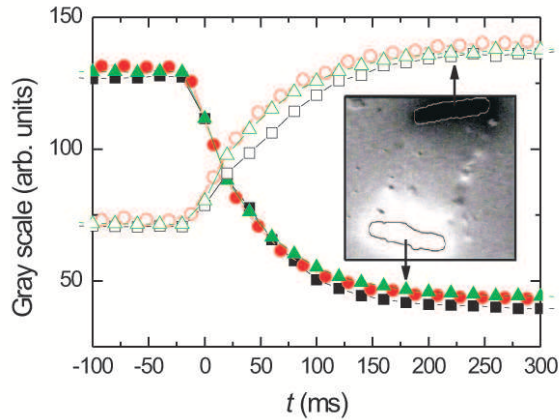}
\caption{(Color online) Time evolution of magneto-optical
intensity at the upper and lower edges of the crystal for the
three cases of \emph{experiment 2}. The different symbols
correspond to different excitation energies: 0.5 $\mu$J (squares),
5 $\mu$J (circles) and 50 $\mu$J (triangles). The inset shows the
regions over which the average intensity was calculated. All
curves fit to an exponential law with a single $\tau=65.8$
ms.\label{Fig.3}}
\end{figure}

Finally, the width of the region over which the simultaneous relaxation of
the spins is expected to occur can be easily estimated using
$l_D=\sqrt{\kappa\tau}$ \cite{Suzuki,QMDTheory}. From the
literature \cite{Macia} it is known that $\kappa$ for Mn$_{12}$ is
in the range of $10^{-5}-10^{-4}$ m$^2$/s hence
$l_D\approx0.8-2.5$ mm. This is of the order of the length of our
sample, which also explains why we have not seen the narrow front
of reversing spins reported in ref. \cite{Suzuki}.

In conclusion we have visualized the magnetic deflagration using
for the first time an optical technique. We have shown that
depending on the environment and the size of the sample, it is
possible to ignite deflagrations with long combustion times and
wide magnetization fronts. We show that a reduction of the
propagation speed of magnetic deflagrations is experimentally
accessible, which opens the possibility of new types of
experiments.

\acknowledgments

This work was supported by FOM (Stichting voor Fundamenteel
Onderzoek der Materie) that is financially supported by NWO
(Nederlandse Organisatie voor Wetenschappelijk Onderzoek) and the
Spanish Ministerio de Educaci\'on y Ciencia (MEyC), contract
MAT2005-06162. A.H.-M. and F.M. thank the MEyC for a research
grant. J.M.H. thanks the MEyC and University of Barcelona for a
Ram\'on y Cajal research contract.

\end{document}